%Paper: hep-th/9307151
%From: CERESOLE@to.infn.it
%Date: Sun, 25 Jul 1993 19:06:01 +0200 (WET-DST)
%Date (revised): Mon, 26 Jul 1993 16:20:09 +0200 (WET-DST)

%%%%%please use TeX and phyzzx
%%macro%%%%%%%%%%%%%%%%%%%%%%%%%%%%%%%%%%%%%%%%%%%%
%%%%%7 figures availables as hard copies from the
%%authors%%%%%%%%%%%%%%%%%%%%%%%
%%%%%     in  the revised version                     %%%%%%%%%%%%%%%%%%%%%%%
%%%%%figure 1 is attached as .ps file at the end of
%%text%%%%%%%%%%%%%%%%%%%%%%%%
%%%%%%%%%%%NO OTHER CHANGE%%%%%%%%%%%%%%%%%%%%%%%%%%%%%%%%%%%%%%%%%%%%%%%%%%%%%
\input phyzzx
\pubtype={ }
\date{ }
%%%%%%%%%%%annamac%%%%%%%
\def\ni{\noindent}
\def\to{\rightarrow}
\def\bar{\overline}
\def\del{\partial}
\def\IR{{\hbox{{\rm I}\kern-.2em\hbox{\rm R}}}}
\def\IC{{\hbox{{\rm I}\kern-.6em\hbox{\rm C}}}}
\def\IP{{\hbox{{\rm I}\kern-.2em\hbox{\rm P}}}}
\def\IZ{{\hbox{{\rm Z}\kern-.4em\hbox{\rm Z}}}}
\def\II{\relax{\rm I\kern-.6em 1}}
\def\IGa{\relax{{\rm I}\kern-.18em \Gamma}}
\def\CP{{\,\IC\IP}}
\def\semi{\mathrel{\raise0.04cm\hbox{${\scriptscriptstyle |\!}$
\hskip-0.175cm}\times}}
\def\ie{\hbox{\it i.e.}}
\def\eg{\hbox{\it e.g.}}
\def\a{\alpha } \def\s{\sigma }
\def\b{\beta }  \def\r{\rho }
\def\c{\gamma } \def\d{\delta }
\def\o{\omega}  \def\l{\lambda}
\def\iy{{\rm y}}

\def\cW{{\cal W} }
\def\cO{{\cal O} }
\def\cN{{\cal N} }
\def\cM{{\cal M} }
\def\cR{{\cal R} }
\def\CY{Calabi--Yau\ }
\def\Kal{K\"ahler\ }
\tolerance=5000
\overfullrule=0pt
\twelvepoint
\def\aff#1#2{\centerline{$^{#1}${\it #2}}}
\titlepage
\line{\hfill POLFIS-TH. 05/93}
\line{\hfill DFTT 34/93}
\title{Duality Group for Calabi--Yau 2--Moduli Space$^\dagger$}
\foot{e-mail address: ceresole@to.infn.it }
\foot{Supported in part by M.U.R.S.T}
\author{A.\ Ceresole$^{\star,1,3}$,
R.\ D'Auria$^{1,3}$ and
T.\ Regge$^{2,3}$}
\vskip2.truecm
\aff1{Dipartimento di Fisica, Politecnico di Torino}
\centerline{\it
 Corso Duca Degli Abruzzi 24, 10129 Torino, Italy}
\line{\hfill}
\aff2{Dipartimento di Fisica Teorica, Universit\`a di Torino}
\centerline{\it
 Via P. Giuria 1, 10125 Torino, Italy}
\line{\hfill}
\aff3{ INFN, Sezione di Torino, Italy.}
\vskip .8 cm
\abstract
{ \ni
We present an efficient method for computing  the duality group $\IGa$
of the moduli space \cM for strings compactified on a \CY manifold described
by a two-moduli deformation of the quintic polynomial immersed in $\CP(4)$,
$\cW={1\over5}(\iy_1^5+\cdots+\iy_5^5)-a\,\iy_4^3 \iy_5^2 -b\, \iy_4^2
\iy_5^3$.
We show that $\IGa$ is given by a $3$--dimensional representation of
a central extension of $B_5$, the braid group on five strands.
}
\vfill
\line{July, 1993\hfill}
\endpage
%%%%%%
%references
%%%%%%%%%
\REF\duality{ K. Kikkawa and M. Yamasaki,  Phys. Lett. {\bf 149B} (1984) 357;
N. Sakai and L. Senda,  Progr. Theor. Phys.  {\bf 75} (1986) 692;
V.P. Nair, A. Shapere, A. Strominger and F. Wilczek,  Nucl.
Phys.  {\bf B287} (1987) 402;
A. Giveon, E. Rabinovici and G. Veneziano,  Nucl. Phys. {\bf B322} (1989) 167;
A. Shapere and F. Wilczek,  Nucl. Phys.  {\bf B320} (1989) 167;
M. Dine, P. Huet and N. Seiberg,  Nucl. Phys.  {\bf B322} (1989) 301;
J. Molera and B. Ovrut,  Phys. Rev.  {\bf D40} (1989) 1150;
J. Lauer, J. Maas and H.P. Nilles,  Phys. Lett.
{\bf B226} (1989) 251 and  Nucl. Phys.  {\bf B351} (1991) 353;
W. Lerche, D. L\"ust and N.P. Warner,   Phys. Lett.
{\bf B231}  (1989) 418;
M. Duff,  Nucl. Phys. {\bf B335}  (1990) 610;
A. Giveon and M. Porrati,  Phys. Lett.  {\bf B246} (1990) 54 and  Nucl. Phys.
{\bf B355} (1991) 422;
Giveon, N. Malkin and E. Rabinovici,  Phys. Lett. {\bf B238} (1990) 57;
J. Erler, D. Jungnickel and H.P. Nilles,   MPI-Ph/91-90;
S. Ferrara, D. L\"ust, A. Shapere and S. Theisen,  Phys.
Lett.  {\bf B233} (1989) 147;
J. Schwarz,  Caltech preprint CALT-68-1581 (1990),
Phys. Lett. {\bf B272} (1991) 239 and  in {\sl Strings: Stony Brook 1991} World
Scientific;
J. Erler, D. Jungnickel and H.P. Nilles,  MPI-Ph/91-81;}
\REF\witte{ E. Witten,  Phys. Lett. {\bf B155} (1985) 151;}
\REF\canetal{ P. Candelas, X.C. de la Ossa, P.S. Green and L. Parkes,
Phys. Lett. {\bf 258B} (1991)~118;
P. Candelas, X.C. de la Ossa, P.S. Green and L. Parkes,
Nucl. Phys. {\bf B359} (1991)~21;}
\REF\onemod{ D. Morrison, {\sl Essays on Mirror manifolds}, S.T.Yau Editor,
Intenational Press (1992); A. Font, Nucl. Phys {\bf B391} (1993) 358;
A. Klemm and S. Theisen,
Nucl Phys. {\bf B389} (1993) 153;}
\REF\cande{P.Candelas, private communication and talk held at ''Eloisatron
Project'', Erice , December 1992; P. Candelas, X. De La Ossa, A. Font,
S. Katz and D. R. Morrison, ''Mirror Symmetry for Two Parameter Models-I'',
CERN-TH.6884/93, in preparation;}
\REF\dafer{R. D'Auria and S. Ferrara, `` String Quantum Symmetries From
Picard-Fuchs Equations And Their Monodromy'', CERN-TH.6777/93 ,
POLFIS-TH.24/93;}
\REF\regpon{T. Regge,  ``The Fundamental Group Of Poincar\'e And The
Analytic Properties Of Feynman Relativistic Amplitudes''  Nobel Symposium
Series VIII (1968); G. Ponzano and T. Regge, Proceedings of the
Varna Conference (1968);}
\REF\prsw{G. Ponzano, T. Regge, E. R. Speer and M. J. Westwater, Commun.
Math. Phys. {\bf 15} (1969) 83 and Commun. Math. Phys. {\bf 18} (1970) 1;}
\REF\zarin{See O. Zariski, ``{\sl Algebraic Surfaces}'', II ed., Springer-
Verlag (1971);}
\REF\zar{O. Zariski,  Ann. for Math {\bf 38} (1937) 131;}
\REF\vankam{E. R. van Kampen,  Amer. J. Math.  {\bf 55} (1933) 255;}
\REF\special{ A. Strominger,  Commun. Math. Phys. {\bf 133} (1990)  163;
L. Castellani, R. D'Auria and S. Ferrara,  Phys. Lett. {\bf B241} (1990) 57;
L. Castellani, R. D'Auria and S. Ferrara,
Class. Quant. Grav. {\bf 1} (1990) 317;}
\REF\cdfll{ A. Ceresole, R. D'Auria, S. Ferrara, W. Lerche and J. Louis,
Int.Jou.Mod.Phys. {\bf A8}  (1993) 79;}
\REF\dkl{ L.J. Dixon, V.S. Kaplunovsky and J. Louis,  Nucl. Phys.
{\bf B329} (1990) 27;}
\REF\cremvan{E. Cremmer and A. Van Proyen, Class. Quantum Grav. {\bf 2} (1985)
 445}
\REF\cafe{ A. Cadavid and S. Ferrara,  Phys. Lett. {\bf B267} (1991) 193;}
\REF\lsw{ W. Lerche, D. Smit and N. Warner,  Nucl. Phys.  {\bf B372} (1992)
87;}
\REF\arnold{V. I. Arnold, S. M. Gusein--Zade and A. N. Varchenko,
 Singularities of Differentiable Maps Vol {bf II} (Birkh\"auser).}
%%%%%%%%%%%%%%%%%%%%%%%%%%%%%%%%%%%%%%%%%%%%%%%%%
%%%%body%%%%%%%%%%%%%%%%%%%%%%%%%%%%%%%%%%%%%%%%%
%%%%%%%%%%%%%%%%%%%%%%%%%%%%%%%%%%%%%%%%%%%%%%%%%
\chapter{Introduction}
Generic string models are believed to exhibit target space duality$^\duality$
,
which is a discrete symmetry acting on the moduli space of the underlying
superconformal field theory (SCFT), or of the relevant compactified
$6$-dimensional manifold. Duality symmetry, whose origin is strictly
related to the fact that the string is a $1$-dimensional object,
leaves invariant the spectrum and supposedly the
interactions, and quite generally describes quantum symmetries of the low
energy effective action.
Such symmetry has been explicitly found in
simple models like toroidal compactifications and their orbifolds, where it
generalizes  the $R\rightarrow 1/R$ symmetry of the bosonic string
compactified on a circle.
The most celebrated example is the effective Lagrangian obtained by
 compactification of the heterotic string on a $6-$torus in the large radius
limit$^ \witte$.
 The \Kal class untwisted modulus corresponding to the volume size
$t=2(R^2+i\sqrt g)$ parametrizes the homogeneous space $SU(1,1)/U(1)$, so
 that this sector of the theory is invariant under $PSL(2,\IR)$.
However, when other (twisted) sectors are introduced and/or quantum corrections
computed, the residual symmetry is given by the duality group $SL(2,\IZ)$ with
generators $t\to t+1$ and $t\to -{1\over t}$.
The determination of the duality symmetry group $\IGa$ for \CY
compactifications, or in broader sense for $(2,2)\  c=9$ SCFT,
is in general a very difficult problem, so far  solved only
for few examples in which the moduli space is one-dimensional. In particular,
in a seminal paper$^\canetal$,
 Candelas et al. treated in detail the class of \CY
 $3$--folds immersed in $\CP(4)$ given by the one parameter deformation of
 the quintic polynomial ${\cW}_{0}$
$$
{\cW} ={\cW} _{0}-\psi\ \iy_{1}\iy_{2} \iy_{3} \iy_{4} \iy_{5}\qquad,\qquad
{\cW}_{0}={1\over5}\left(\iy_{1}^{5}+\iy_{2}^{5}+\iy_{3}^{5}+\iy_{4}^{5}
+\iy_{5}^{5}\right)
\eqn\can
$$
where $\iy_i$ are homogeneous coordinates on $\CP(4)$.
Using the monodromy properties of the solutions of
the Picard-Fuchs equations, these authors were able to reconstruct the full
duality group of the $1$--dimensional space parametrized by $\psi$.
It was found that $\IGa$ has two generators, $\{A,T\}$ of order $5$ and
$\infty$ respectively, and their representation in terms of
$4\times 4$--matrices acting on the periods was given. After that, several
other
examples of $1$--dimensional deformations of other \CY manifolds
defined by polynomials in weighted projective space have been developed
along the same lines$^\onemod$.

The general computation of the duality group for more than
one modulus  has not been successfully attempted so far, because of the
mathematical
complexity that is encountered in going from the $1$--dimensional case to the
higher dimensional ones. However, a $2$--moduli case is presently under
investigation$^{\cande}$.
Moreover, it has been observed$^{\dafer}$ that for any number
of moduli, the translational symmetry on the special variables
$t^a\to t^a+n^a\, , n^a\in \IZ$, is always a subgroup of the duality
group.

In this paper we present an efficient method for  determining the
duality group for  a $2$--moduli deformation of the quintic
polynomial ${\cW}_0$,
$$
{\cW} = {\cW}_0 - a\, \iy_{4}^3 \iy_{5}^2-b\, \iy_{4}^3 \iy_{5}^2\ ,
\eqn\pot
$$
This example of \CY manifold is a subspace of the general $101$--dimensional
deformation of ${\cW}_0$ which gives rise to zero Yukawa couplings for the
associated effective Lagrangian. Because of that, if taken by itself, it
constitutes more of a toy model as far as  the low energy Lagrangian is
concerned.
However, it provides a $2$-moduli example for which  the duality
group can be determined completely, displaying the power of
some general techniques of algebraic geometry which
were developed and applied to the study of the monodromy groups of Feynman
integrals many years ago$^{\regpon,\prsw}$ .

Our result for the duality group is surprisingly simple: $\IGa$ is given
by an $U(1,2)$ valued (projective) representation of
 $B_5$,  the braid group on five strands. The representation
 acts on the periods associated to the uniquely defined holomorphic
three--form $\Omega$ which always exists on a \CY manifold.
 In terms of the defining polynomial $\cW$, the fundamental period is
 defined by the integral representation$^{\cafe,\lsw}$
$$
\o_0(a,b)=\oint_\Gamma{\o\over{\cW (\iy;a,b)}}\ ,
\eqn\period
$$
where $\o$ is the volume element
$$
\o=\sum(-1)^i \iy_i\, d\iy_1\wedge\cdots\wedge\hat{d\iy_i}\wedge\cdots
\wedge d\iy_5
\eqn\volume
$$
 (the hat means that the corresponding differential must be omitted),
and $\Gamma$ is an element of the basis for the homology cicles of
$H_{(4)}(\CP(4)-\cW ;\IZ)$. There are as many independent integrals
$\o_0^I$ as there are elements of the basis, $\Gamma^I\subset H_{(4)}
(\CP(4)-\cW;\IZ)$.

Quite generally, if $L^{N-1}$ is the singularity locus of an algebraic curve
$\cW$ parametrized by $N$ moduli, the monodromy group $\IGa$ acting on the
 periods of $\cW$ is given by a representation of the fundamental group $\pi_1$
of the embedding space $\CP(N)$.

The computation of the homotopy group $\pi_1$ is
 based on the use of the following two theorems$^{\zarin , \zar}$.

{\bf Theorem 1} (Picard-Severi). {\sl Let $L^{N-1}$ be the $N-1$
 complex dimensional singular locus of a given algebraic curve.
 If the (complex) projective line
 $\CP(1)\subset \CP(N)$ is generic with respect to $L^{N-1}$ (\ie, it avoids
all singular points of $L^{N-1}$), then we have the isomorphism
$$
\pi_1 \left(\CP(1)-(\CP(1)\cap L^{N-1});B\right)/G \approx
\pi_1\left(\CP(N)-L^{N-1};B\right)
\eqn\picsev
$$
where $B$ is the base point and $G$ is an invariant subgroup
of
$\pi_1\left(\CP(1)-\right.$ $(\left.\CP(1)\cap L^{N-1});B\right)$. }

In other words,
$\pi_1\left(\CP(N)-L^{N-1};B\right)$ is obtained from
$\pi_1\left(\CP(1)-\right.$ $\left.
(\CP(1)\cap L^{N-1});B\right)$ by adding the identities
$\gamma=\II\ \  \forall \gamma \in G$.

A method for deriving such identities has been provided by Van Kampen$^{
 \vankam}$,  and we shall use it in our particular case to
obtain the isomorphism of eq. \picsev .
As the singular locus of the algebraic curve $\cW$ is given by an equation
of the form $L(a,b)=0$, we are interested in the case $N=2$.

The second theorem, due to Zariski$^{\zar}$,  allows  to understand
that, as far as the identification of the fundamental group $\pi_1$ is
concerned, the case of more than two moduli can be essentially reduced to the
$N=2$ case, so that the general computation of
$\pi_1$  in  will not be more difficult  than  the one
 under study.  However,  for several variables it is in general much harder
 to find the
 singular locus and the behaviour of the algebraic
 curve in its neighbourhood, and therefore the determination of the monodromy
group can be more involved.

{\bf Theorem 2} (Zariski). {\sl If the complex projective plane $\CP(2)$ is
generic with respect to $L^{N-1}$ and if
 $B\in \left(\CP(2)-\CP(2)\cap L^{N-1}\right)$, then the map
$$
\pi_1\left(\CP(2)-\CP(2)\cap L^{N-1};B\right)\rightarrow \pi_1\left(
\CP(N)-L^{(N-1)};B\right)
\eqn\zar
$$
is an isomorphism.}

We see that, in virtue of Zariski theorem,  the study of the
homotopy group of $\CP(N)-L^{N-1}$ is reduced  to the study of the homotopy
 of the complement of the $1$--dimensional curve $L^1=\CP(2)\cap
L^{N-1}$ on a generic two-dimensional section. Since the
 singular locus of the curve \pot\  is already one-dimensional
 ( $N=2$ ),  theorem 1. is sufficient for our present purposes.

Before proceeding to the actual determination of the duality group,
we recall what is the local geometry associated to the moduli
space $\cM$ of $\cW$. The two-moduli family of \CY deformations
given by eq. \pot\  was first introduced in ref. \dkl\  as a
particular tensor product example of five copies of minimal models
with $k=3$ and $c=9/5$. It was observed that in such model there are
restrictions given by charge conservation which enforce the condition
$W_{\a\b\c}=0$ for the Yukawa couplings. Because of that, the constraint
of Special Geometry$^{\special}$
$$
R_{\a\bar\b\c\bar\d}=g_{\a\bar\b}g_{\c\bar\d}
+g_{\a\bar\d}g_{\c\bar\b}-e^{2 K} W_{\a\c\rho}
W_{\bar\b\bar\d\bar\sigma} g^{\rho\bar\sigma}
\eqn\specgeo
$$
reduces to
$$
R_{\a\bar\b\c\bar\d}=g_{\a\bar\b}g_{\c\bar\d}
+g_{\a\bar\d}g_{\c\bar\b}\qquad ,\qquad \a ,\bar \b ,
\c ,\bar\d =1,2\, \ .
\eqn\rest
$$
Thus, the local geometry of  $\cM$ is  given by a $2$--dimensional
 \Kal manifold with constant curvature, which according to
the classification of ref. \cremvan\  ,
corresponds to the coset space ${U(1,2)\over{U(1)\otimes U(2)}}$.
The global structure of such moduli space is determined by modding out
the discrete isometries given by the duality group $\IGa$.

In principle, for the $2$--moduli case, one would expect a $6$--dimensional
representation of the modular group. In fact,
the dimension of the $H^{(3)}$ cohomology group is given by $2 h_{21}+2$,
where $h_{21}={\sl dim}\,H^{(2,1)}$,  the
number of complex structure moduli of the \CY manifold. It turns
out, however, that our representation is only $3$--dimensional, as this
specific model is singular due to the vanishing of the Yukawa couplings
$W_{\a\b\c}$.
In such case, the Picard-Fuchs equations
for the periods of $\cW$ are of second order rather than fourth-order$^{\cdfll}
$.
The $6$--dimensional  representation, which in the symplectic
basis is valued in $Sp(6,\IZ)$,
splits into two $3$--dimensional representations of $U(1,2)$ according
to the embedding $6=3+\bar 3$ of $U(1,2)$ in $Sp(6)$. Later, we shall verify
explicitly  that the $3$--dimensional representation of
$B_5$ is actually valued in $U(1,2)$.

\chapter{The Fundamental Group of\  $\cW (\iy;a,b)$.}
In this section we compute the fundamental group $\pi_1 (CP(2)-L;B)$ of the
algebraic curve $\cW=0$.
The first step consists of finding the singular locus $L$
of eq. \pot , which
is given by solving simultaneosly the equations
$$
{{\del\cW}\over{\del \iy_{i}}}=0\qquad i=1,\cdots,5\, .
\eqn\risu
$$
A straightforward computation yields the $1$--dimensional complex curve
$$
L(a,b)=108 (a^5+b^5)-80 a^3 b^3-165 a^2 b^2+30 a b-1 =0
\eqn\locus
$$
which represents the locus of the complex points of the original curve
${\cW} = 0 $ where two or more of the roots coincide.

For the derivation of the Van Kampen relations among the homotopy generators
around the various branches of $L(a,b)$, it is important to know where $L(a,b)$
itself has multiple points. These are found by solving the equations
$L(a,b)={{\del L}\over{\del a}}={{\del L}\over{\del b}}=0$, which give the
location of the multiple roots
$$\eqalign{
(a,b)&=(\r^k\,, \r^{-k})\cr (a,b)&=-{1\over4}(\r^k\,, \r^{-k})\qquad
k=0,\cdots,4\cr}
\eqn\roots
$$
where $\r=e^{2\pi i/5}$. The first  set of roots  in \roots\
 corresponds to
nodes with two distinct complex conjugate tangents (which are isolated points
for the real section of
\locus\  represented by real values of $a$ and $b$).
 The second set instead represents (second order) cusps,
since the Hessian ${{\del^2 L}\over{\del a \del b}}$ is degenerate at these
points.

We may obtain a more elegant and geometrically intuitive representation of the
curve $L(a,b)=0$ by choosing new coordinates $(p,q)$  such  that the
real section corresponding to real values of $p$ and $q$ exhibits the previous
singular points in the real $(p,q)$ plane. It is sufficient to set
$$\cases{
a&= p + i q\cr
b&= p -- i q\cr}
\eqn\change
$$
and we find
$$
\eqalign{L(p,q)= &-1+30(p^2+q^2)-165 (p^4+q^4)-80(p^6+q^6)
+216 p^5\cr\ &-330p^2q^2 -2160p^3q^2-240 (p^4q^2+p^2 q^4) +1080 p q^4\, .\cr}
\eqn\new
$$
In the real plane of $(p,q)$ the multiple points \roots\  take now the real
values
$$
\eqalign{
(p,q) &=(\cos{{2 \pi k}\over 5},\sin{{2 \pi k}\over5})\cr
(p,q) &=-{1\over4}(\cos{{2 \pi k}\over 5},\sin{{2 \pi k}\over5})\cr}
\eqn\newroots
$$
respectively. Actually, the curve $L(p,q)=0$ can be put in a parametric
form by setting
$$
\left\{\eqalign{p &={1\over{5}}(3 \cos2 t+2 \cos 3 t)\cr
       q &={1\over{5}}(3 \sin 2 t- 2 \sin 3 t)\qquad 0\leq t\leq2\pi\cr}
\right .\eqn\ipo
$$
and can be recognised as a pentacuspidal hypocycloid (the curve described
by a point of a circle of radius $R=1/5$ rolling inside a circle
of $R=1$), whose graph is represented in fig. 1.

According to Picard-Severi theorem, we now take a generic line $C$ through
the base point $B=(0,0)$ which intersects the real branch of the hypocycloid
in four (finite) points $P_i$ (fig. 2).
 To each of such representative points, we attach a
generator of $\pi_1(\IC-\IC\cap L;B)$ by constructing a loop which leaves $B$
along a straight line, makes an infinitesimal loop around $P_i$
counterclockwise in the complex plane $\IC$ containing the real line, and comes
back to $B$ again in the opposite direction.
 If the straight line encounters
a real branch of $L$, it will be taken to undercross it by describing a
 small semicircle in the complex plane. If by varying the angular
 coefficient of the straight line no critical point of $L$ is encountered,
 the corresponding points give rise to equivalent loops. Generally
inequivalent loops are obtained if two straight lines intersect $L$ along
points belonging to two different real branches of $L$ emanating from the
critical points. We thus obtain $15$ loops, $5$ of which go around the sides
of the pentagonal figure described by $L$ and $10$ going around the branches
emanating from the cusps (5 of them are shown in fig. 3.).

We now quote the Van Kampen relations$^{\vankam}$
 between loops which, added to the free
group generated by the above mentioned $15$ generators, make it isomorphic
to $\pi_1(\CP(2)-L;B)$.

For the nodes corresponding to transversal intersections of the real branches
of $L$, we have (see fig. 4)
$$
\a\b=\b\a\qquad ; \qquad\a=\a'\qquad ;\qquad\b=\b'
\eqn\tran
$$
\ie, we can slide the representative loops across the node without any
change, and the loops around two branches commute. In this way, the $15$
generators reduce to five independent ones

For each cusp we have the relation (see fig. 5)
$$
\a\b\a=\b\a\b
\eqn\aba
$$
Let us enumerate in increasing order the five branches of the hypocycloid
described successively by the point of the small circle rolling inside the
big circle (fig. 2).
 Then,  denoting by $\a_i\, ,i=1,\cdots ,5$ the loops winding
around the five branches of fig. 1, we have the following set of relations
among the generators
$$
\eqalign{
\a_1\a_3 = \a_3\a_1 &\qquad \a_1\a_2\a_1 =\a_2\a_1\a_2 \cr
\a_1\a_4 = \a_4\a_1 &\qquad \a_2\a_3\a_2 =\a_3\a_2\a_3 \cr
\a_2\a_4 =\a_4\a_2  &\qquad \a_3\a_4\a_3 =\a_4\a_3\a_4 \cr
\a_2\a_5 =\a_5\a_2  &\qquad \a_4\a_5\a_4 =\a_5\a_4\a_5 \cr
\a_3\a_5 =\a_5\a_3  &\qquad \a_5\a_1\a_5 =\a_1\a_5\a_1 \cr}
\eqn\rela
$$
\ni We note that the subset of relations \rela\ not involving $\a_5$
coincide with the defining relations of the
braid group $B_5$, with four generators $\a_i\,,i=1,\ldots,4$,
$$
\left\{\eqalign{
\a_i \a_j&=\a_j \a_i\qquad|i-j|\ge2\cr
\a_i \a_{i+1} \a_i &= \a_{i+1} \a_i \a_{i+1}\qquad i=1,\ldots,3\cr}\right .
\eqn\trecce
$$
 However, it may be easily verified that the element of
$B_5$ exchanging the first and the fifth strand can be written in terms
of the four generators $\a_i$ by the word
$$
\a_5=( \a_4 \a_3 \a_2 ) \a_1 ( \a_4 \a_3 \a_2 )^{-1}
\eqn\bifive
$$
With a little effort, using the Van Kampen relations for $\a_1,\cdots,\a_4$
one checks that the extra relations of \rela\ involving $\a_5$ are indeed
verified.
Therefore, we come to the conclusion that, if the relation \bifive\
also holds among the monodromy generators $\a_i\, (i=1,\cdots,5)$ of the
hypocicloyd, then $\pi_1(C(2)-L;B)$ is isomorphic to $B_5$.

The reason why we do not find eq. \bifive\ among the Van Kampen relations
is  that we have not considered the critical point of $L$
at $\infty$ and the associated generator. Rather than studying such critical
point, we shall give evidence in the sequel that eq. \bifive\  must be
satisfied by the monodromy generators, so that indeed $B_5$ coincides with
the fundamental  group associated to $\cW$.

\chapter{Behaviour of the Periods Around the Singular Curve.}

Till now we have only considered the abstract presentation of the
fundamental group in terms of its generators $\a_i$.
 To obtain an explicit realization on the periods of $\cW$, it is
necessary to consider their leading behaviour  in the
neighbourhood of the singularity locus $L(a,b)$.
Let us evaluate the integral defined in \period\ on a suitable contour.
 Setting $\iy_3=1$, we may rewrite it in the following way
$$
\o_0=5\oint d\iy_2 d\iy_4 d\iy_5
 \oint{{d \iy_1}\over{\iy_1^5+f(a,b,\iy_2,\iy_4,\iy_5)}}\ ,
\eqn\newint
$$
where
$$
f=\iy_2^5+\iy_4^5+\iy_5^5+1-5 a\, \iy_4^3\iy_5^2-5 b\, \iy_4^2\iy_5^3\ .
\eqn\effe
$$
Performing the last integration on the cycle $|\iy_1 |={\it const}$
 (posing  $\iy_1^5=t$), gives immediately
 ${{2\pi i}\over5}(-f)^{4/5}$, so that we may write
$$
\eqalign{
\o_0 &= 2\pi i
\oint {{d\iy_2 d\iy_4 d\iy_5}\over{(
\iy_2^5+\iy_4^5+\iy_5^5+1-5 a\, \iy_4^3\iy_5^2-5 b\, \iy_4^2\iy_5^3)^{4/5}}}\cr
  \   &=2\pi i\oint d\iy_4 d\iy_5
\oint{{d \iy_2}\over{[\iy_2^5+g(a,b,\iy_4,\iy_5)]^{4/5}}}\cr}
\eqn\ointegra
$$
where
$$
g=\iy_4^5+\iy_5^5+1-5a \iy_4^3\iy_5^2-5 b\iy_4^2\iy_5^3\ .
\eqn\gigi
$$
Again, with $\iy_2=g^{1/5} u$, the last integral on the cycle
 $|\iy_2|={\it const}$ yields
$$
g^{-3/5}\oint{{du}\over{(u^5+1)^{4/5}}}
\eqn\uinte
$$
which is a number independent of $a$ and $b$, so that we arrive to
$$
\o_0={\it const}\times\oint{{d\iy_4 d\iy_5}\over{g^{3/5}}}
\eqn\quasi
$$
After the further change of variables
$$
\iy_4=\s\tau^{1/2}\qquad\, ;\qquad\iy_5=\s\tau^{-1/2}
\eqn\cambio
$$
 we find
$$
\o_0={\it const}\times
\oint { {d\tau}\over\tau}
\oint { {\s d\s}\over{[1+\s^5 h(\tau)]^{3/5}} }
\eqn\pitau
$$
where $h(\tau)=\tau^{5/2}+\tau^{-5/2}-5 a\tau^{1/2}-5 b\tau^{-1/2}$ .
Setting at last $\xi=\s h^{1/5}$, the $\s$ cyclic integral gives $h^{2/5}$
times a purely numerical integral. Therefore we obtain that the
periods $\o_0$ can be written as a simple $1$--dimensional integral
$$
\o_{ij}={\it const}\times
\oint_{\c_{ij}} { {d\tau}\over{(\tau^5-5 a\, \tau^3-5 b\,\tau^2+1)^{2/5}} }
\eqn\periodo
$$
\ni along a set of suitably chosen contours $\c_{ij}$.
The quintic polynomial $P(\tau)=\tau^5-5 a \tau^3-5 b\tau^2+1 $ has five roots
$\tau_i$, and the possible singularities of the integral \periodo\ arise as
pinching singularities due to the coincidence of two roots encircled by a
figure-eight contour (see fig. 6.).

One can check that the singular locus of the singularities of $P(\tau)$
is given again by the curve $L(a,b)=0$ of eq. \locus\ by computing the
resultant
between $P(\tau)$ and ${{dP}\over{d\tau}}$.

At this point, we may conclude that the monodromy group is exactly $B_5$.
In fact we  note that the braid group $B_n$ can be identified with the
fundamental group of the space of all unordered sets of $n$ distinct complex
numbers $\tau_i\ i=1\ldots,n$. More precisely,
$B_n=\pi_1(\IC^n/S_n\backslash S/S_n)$, where $S_n$ is the permutation group on
five elements and $S$ is the union of the hyperplanes $\tau_i=\tau_j\
 \forall (i,j)^
{\arnold}$ . Identifying the numbers $\tau_i$ with the roots of a polynomial
$P_n(\tau)$ we see that in our case the monodromy group of the periods $\o_0$
must be a subgroup of $B_5$, since $P(\tau)$ is a quintic polynomial. On the
other hand, $B_5$ contains the element given by \bifive\ exchanging the
strands $1$ and $5$. Since
$$
\a_5=\a_4\a_3\a_2\a_1\a_2^{-1}\a_3^{-1}\a_4^{-1}
\eqn\afive
$$
satisfyes the relations \rela\ involving $\a_5$, the group whose presentation
is given by \rela\ must actually coincide with $B_5$.

There is yet another independent argument leading to the same conclusions.
Given a general polynomial of $5$--th degree, we may always fix $3$ of its
$5$ coefficients by a M\"obius transformation to arbitrary values. As
$P(\tau)$ contains only two parameters, it is infact a gauge fixed form of
a generic quintic polynomial, and therefore the associated subgroup of $B_5$
is really $B_5$ itself.

One can show that the
set of all possible figure eight contours $\c_{ij}$ encircling a couple of
 roots can be
expressed linearly in terms of three of them, thus confirming that the number
of linearly independent periods of $\cW$ is indeed equal to three. To see this,
let us denote by $\c_{i,i+1}$ the figure--eight contour encircling two
 consecutive roots $\tau_i ,\tau_{i+1}$ (loops encircling non consecutive roots
are easily written as products of the $\c_{i,i+1}$'s, \eg
$\c_{i,i+2}=\c_{i,i+1}
\circ \c_{i+1,i+2}$, etc.), and by $\o_{i,i+1}$ the corresponding period. Only
three of them are indipendent, since they satisfy the following two relations
$$
\eqalign{
\sum_{k=1}^5 & \o_{k,k+1}=0\cr
\sum_{k=1}^5 & e^{i(k-1)4\pi/5} \o_{k,k+1}=0\cr}
\eqn\duere
$$
\ni The first relation easily follows from the fact that $\c_{1,2}\circ
\c_{2,3}\circ \c_{3,4}\circ\c_{4,5}\circ\c_{5,1}$ is homotopic to a single
loop encircling all the five roots of $P(\tau)$, and the integral is regular
at $\infty$. The second relation can be obtained by observing that
$$
\o_{i,i+1}=I_{i+1}-e^{-4\pi i/5} I_i
\eqn\trere
$$
where $I_i$ is the integral around a loop winding counterclockwise
 around the simple
root $\tau_i$. Therefore, the representation of $B_5$ on the $\o_{i,i+1}$
is $3$--dimensional, thus confirming the observation made in the introduction
that the vanishing of the Yukawa couplings reduces the $6$--dimensional
$Sp(6,\IZ)$--valued representation of the periods into the $3+{\bar 3}$-
representation which will be later shown to belong to  $U(1,2)$.

The behaviour of $\o_{ij}(a,b)$ around the $L(a,b)=0$ singularity can be now
obtained by expanding $P(\tau)$ in the neighbourhood of a point $\tau_0$
where $P(\tau)$ vanishes together with its first derivative:
$$
\eqalign{
&P(\tau)=(\tau-\tau_0)^2+H(\tau,a,b)\cr
&\lim_{\tau\to\tau_0} H(\tau,a,b)=L(a,b)\cr}
\eqn\limite
$$
If we now set $\tau-\tau_0=H^{1/2}\eta$, then we find
$$
\o_{ij}={\it const}\, L^{1/10} \ \oint{{d\eta}\over{(\eta^2+1)^{3/5}}}\equiv
{\it const} \,  L^{1/10} (a,b)
\eqn\behav
$$
thus finding that, upon performing a loop around a branch of the hypocycloid,
the integral acquires a phase $e^{i\pi/5}\equiv z$.

An alternative way of reaching the same conclusion is to consider the effect of
analytic continuation of $\o_{i,i+1}$ around the loop $\a_i\in B_5$ in the
$(a,b)$--plane,
$$
\o_{i,i+1}\buildrel\a_i\over\to e^{i\pi/5}\o_{i,i+1}
\eqn\azione
$$
Indeed, when the path $\a_i$ winds around the $i$--th branch of the
hypocycloid,
the roots $\tau_i(a,b),\tau_{i+1}(a,b)$ of the polynomial $P(\tau)$ are
exchanged. In the $\tau$ plane, $\c_{i,i+1}$ gets deformed as in fig.7.
The final result is a new circuit followed in opposite direction where the
base point $B$ is on a different Riemann sheet. This gives the same integral
as before except for a phase $-(e^{2\pi i})^{-2/5}= z$.

The correctness of this result can also be ascertained by studying the
leading behaviour of the period $\o_0$ from the Picard-Fuchs equations
obeyed by it.
Using the methods of \cafe--\lsw , one derives the following set
of three partial differential equations
$$
\eqalign{
{{\del^2\o_0}\over{\del a^2}}
&
={1\over{1-4 ab}}\left[6 b^2{{\del^2\o_0}\over{\del a\del b}}
+3 a{{\del^2\o_0}\over{\del b^2}}
+12 b{{1-2 a b}\over{1-4 a b}}{{\del\o_0}\over{\del a}}
+{{12 a^2}\over{1-4a b}}{{\del\o_0}\over{\del b}}\right]
\cr
{{\del^2\o_0}\over{\del b^2}}
&
={1\over{1-4 ab}}\left[6 a^2{{\del^2\o_0}\over{\del a\del b}}
+3 b{{\del^2\o_0}\over{\del a^2}}
+12 a{{1-2 a b}\over{1-4 a b}}{{\del\o_0}\over{\del b}}
+{{12 b^2}\over{1-4a b}}{{\del\o_0}\over{\del a}}\right]
\cr
{{\del^2\o_0}\over{\del a\del b}}
&
={1\over{1-13 ab}}\left[6 a^2{{\del^2\o_0}\over{\del a^2}}
+6 b^2{{\del^2\o_0}\over{\del b^2}}
+11 a{{\del\o_0}\over{\del a}}+11 b{{\del\o_0}\over{\del b}}+\o_0\right]
\cr}
\eqn\fuchspic
$$
The study of the singularities of eqs. \fuchspic\ is better achieved by writing
the associated linear system. Setting
$$
\eqalign{
\o_1&={{\del\o_0}\over{\del a}}=\oint_{\Gamma}{{\iy_4^3 \iy_5^2}\over{\cW^2}}
\,\o\cr
\o_2&={{\del\o_0}\over{\del b}}=\oint_{\Gamma}{{\iy_4^2 \iy_5^3}\over{\cW^2}}
\,\o\cr}
\eqn\periods
$$
and by elimination of the mixed derivatives in \fuchspic\  we find
$$
{\del\over{\del a}}{\bf \Pi}=A(a,b){\bf \Pi}\qquad ,\qquad
{\del\over{\del b}}{\bf \Pi}=B(a,b){\bf \Pi}\qquad ,\qquad
{\bf \Pi}=\pmatrix{\o_0\cr\o_1\cr\o_2\cr}
\eqn\linear
$$
where the $3\times3$ matrices $A(a,b)$ and $B(a,b)$ are rational functions
of $a$ and $b$ whose denominator contains the singular locus $L(a,b)$.
Actually, most of the matrix elements of $A$ and $B$ also contain an extra
factor of $(1-4 ab)$ whose appearance would imply the extra singular locus
$4 a b=1$. However, such singularity is a gauge artifact, as
the two systems in \linear\ are  covariant under gauge transformations
$$
A\to {\cN}^{-1}A{\cN}-{\cN}^{-1}\del{\cN}
\eqn\gauge
$$
(and similarly for $B$), where $\cN$ belongs to the Borel subgroup of
lower triangular matrices of $GL(3,\IC)$. The exam of the Picard-Fuchs system
in the neighbourhood of the variable $v=ab=1/4$ gives in fact perfectly regular
solutions in the new basis
$$
\eqalign{
\tilde{\o_0}&\equiv\o_0\cr
\tilde{\o_1}&=\o_1+8 a^5\o_2\cr
\tilde{\o_2}&=2 (64 a^5-5)\o_1-(1+32 a^5)\o_2\cr}
\eqn\newva
$$

The same conclusion is found by replacing the two linear systems in \linear\
by two $3$rd order ordinary differential equations in $a$ and $b$ for $\o_0$,
 with coefficients depending on the other variable, where the singularity
$ab=1/4$ is absent. Furthermore, the Fuchsian analysis of the linear system
\linear\ or of the two $3$rd order differential equations in the neighbourhood
of $L(a,b)$=0 gives again the behaviour $L^{1/10}$ around the singularity,
thus confirming our previous analysis.

\chapter{The monodromy generators}

In this section we determine the representation $\cR(\a_i)\subset GL(3,\IC)$
of the fundamental group acting on $V$, the vector space spanned by
three independent periods. The computation is made by extending the
representation of the group $B_5$ to a representation on the group ring
over the complex field $\IC$. Noting that  $\a_i$ acts on the period
$\o_{i,i+1}$ as an analytic continuation around the $i$--th branch of
$L(a,b)$, the corresponding discontinuities are given by
$$
\eqalign{
(\cR(\a_i)-1)\o_{i,i+1} &=(z-1)\o_{i,i+1}\cr
(\cR(\a_i)-1)\o_{j,j+1} &=0\qquad\qquad i\neq j,j+1\cr}\ .
\eqn\disconti
$$
Hence we introduce the (Picard--Lefschetz) discontinuity operators$^{\arnold}$
$$
\cR(\a_i)=(z-1)u_i+1
\eqn\disco
$$
where $u_i$ are $1$-dimensional projection operators obeying $u_i^2=u_i$.
 The first set of relations \rela\
imply
$$
u_i u_j=u_j u_i
\eqn\ucom
$$
where $i,j$ are non contiguous indices and we are considering $1,2,3,4,5$
cyclicly ordered so that $1$ and $5$ are contiguous. By right multiplication
with $u_j$ and left multiplication with $u_i$ we get
$$
\eqalign{
u_i u_j & = u_j u_i u_j\equiv {\l}\, u_j\cr
u_i u_j & = u_i u_j u_i\equiv {\l}\, u_i\cr}
\eqn\unidi
$$
where we have used the fact that since the $u_i$ are $1$-dimensional projection
operators, for any operator $\cO$,
$$
u_i {\cO} u_i ={\l}_{\cO}\, u_i
\eqn\calo
$$
Eqs. \unidi\  then imply
$$
u_i u_j =u_j u_i =0
\eqn\zero
$$
which is a relation much stronger than \ucom\ . Indeed, eq. \zero\ has an
 intuitive meaning since ${\it e.g.}\ u_1 ,u_3$ correspond to projection of
the integral \periodo\ around the disconnected figure eight circuits
winding the roots $\tau_1,\tau_2$ and $\tau_3,\tau_4$ respectively.
{}From the second set of \rela\ we get
$$
(z-1)^2 u_i u_{i+1} u_i +z u_i=(z-1)^2 u_{i+1} u_i u_{i+1}+z u_{i+1}
\eqn\prima
$$
Again, since the $u_i$ are $1$-dimensional projection operators, we have
$$
\eqalign{
u_i u_{i+1} u_i &={\r}\, u_i\cr
u_{i+1} u_i u_{i+1} &={\s}\, u_{i+1}\cr}
\eqn\seconda
$$
Multiplying the two equations in \seconda\ by $u_{i+1}$ on the right
  and $u_i$ on the left respectively, we find $\r=\s$,
so that \prima\ gives
$$
u_i ((z-1)^2\r+z )=u_{i+1} ((z-1)^2\r+z)
\eqn\terza
$$
or
$$
\r=-{z\over{(z-1)^2}}
\eqn\quarta
$$
Using the relation \afive\  in the form
 $\a_5 \a_1 \a_3 \a_2= \a_4 \a_3 \a_2 \a_1$,  we find from \disco\
$$
\eqalign{
&(z-1)^3 u_5 u_4 u_3 u_2+(z-1)^2 u_5 u_4 u_3 +(z-1) u_5 u_4 + u_5 = \cr
&=(z-1)^3 u_4 u_3 u_2 u_1+(z-1)^2 u_3 u_2 u_1 +(z-1) u_2 u_1 +u_1\cr}
\eqn\quinta
$$
By right multiplication with $u_2$ and left multiplication with $u_5$,
using \seconda\ -- \quarta\  we get
$$
(z-1) z^2 u_5 u_4 u_3 u_2 = u_5 u_1 u_2
\eqn\sesta
$$
where the final form of the coefficient on the l.h.s. of \sesta\ is
obtained by using the relations obeyed by the tenth roots of unity
$z=e^{i\pi/5}$ (\eg\   $1-z+z^2-z^3+z^4\equiv0$).
Obviously, equations analogous to \sesta\ are also obeyed by similar products
of $u_i$'s with indices cyclically permuted. Such relations allow us
to replace a product of four contiguous $u_i$ operators in decreasing order
with ( a coefficient times ) the product of three contiguous $u_i$'s in
increasing order, the first and the last factors being the same in
both expressions.

We are now ready to construct the explicit representation for the $\a_i$'s.
We select three arbitrary linearly independent basis vectors $\o_{51},
 \o_{12}
, \o_{45}$ defined as the eigenvectors of $u_5, u_1 , u_4$ corresponding
to eigenvalue one
$$
\eqalign{
u_5\, \o_{51} &=\o_{51}\equiv\Psi_{51}\cr
u_1\, \o_{51} &={1\over{z-1}}\,\o_{12}\equiv\Psi_{12}\cr
u_4\, \o_{51} &={{1-z}\over {z}}\,\o_{45}\equiv\Psi_{45}\cr}
\eqn\sette
$$
where the factors in front of $\o_{ij}$ have been chosen in such a way
that the action of the cyclic permutation $Z$ of $\IZ_5\subset B_5$
gives $Z \o_{i,i+1}=\o_{i+1,i+2}$ with no extra phase.
The application of $u_i\, ,i=1,\ldots,5$ to any basis vector gives a linear
combination of them, namely
$$
u_i u_k \Psi_{51}=p_i\Psi_{51}+q_i\Psi_{12}+r_i\Psi_{45}\qquad k=5,1,4
\eqn\otto
$$
It is now easy to compute the coefficients $p_i,q_i,r_i$ by repeated
use of the formulae \seconda -\quinta. For instance, if we take $i=3$,
then
$$
\eqalign{
u_3\Psi_{51}&=u_3 u_5 \Psi_{51}=0\cr
u_3\Psi_{45}&=u_3 u_4\Psi_{51}=p\Psi_{51}+q u_1 \Psi_{51}+r u_4\Psi_{51}\cr}
\eqn\nove
$$
On the other hand,
$$
u_3 u_4 \Psi_{51}=u_3 u_4 u_5 \Psi_{51}
\eqn\dieci
$$
and using \sesta
$$
u_3 u_4 u_5\Psi_{51}=(z-1) z^2 u_3 u_2 u_1 u_5\Psi_{51}=p\Psi_{51}
+q u_1\Psi_{51}+r u_4 \Psi_{51}
\eqn\undici
$$
Multiplying the last two sides by $u_2$ on the left  we obtain
$$
\eqalign{
&-(z-1) z^2{z\over{(z-1)^2}} u_2 u_1\Psi_{51}=q u_2 u_1\Psi_{51}\cr
&\to q=-{{z^3}\over{z-1}}\cr}
\eqn\dodici
$$
Applying now $u_3$ on the left of \nove\ , we find
$$
u_3 u_4 \Psi_{51}= r\, u_3 u_4 \Psi_{51}\qquad\to {r=1}
\eqn\tredici
$$
Finally, acting with $u_4$ in \nove\ we have
$$
\eqalign{
&u_4 u_3 u_4 \Psi_{51}=(p+r) u_4 \Psi_{51}\cr
&\to -{z\over{(z-1)^2}} u_4 \Psi_{51}=(p+r)u_4\Psi_{51}\qquad\to
p={{1-z^3}\over{(z-1)^3}}\cr}
\eqn\quattor
$$

In the same way one can compute the coefficients for the action of any other
projection operator $u_i$. The final result for the monodromy operator $\a_i$
on the basis $\{\o_{45},\o_{51},\o_{12}\}$ is
$$
\eqalign{
\cR(\a_1)=\pmatrix{1&0&0\cr0&1&1\cr0&0&z\cr}\qquad &,\qquad
\cR(\a_2)=\pmatrix{1&0&0\cr0&1&0\cr z^4&-1+z-z^2&z\cr}\cr
\cR(\a_3)=\pmatrix{z&z^2(1-z)&z^2\cr0&1&0\cr0&0&1\cr}\qquad &,\qquad
\cR(\a_4)=\pmatrix{z&0&0\cr-z&1&0\cr0&0&1\cr}\cr
\cR(\a_5)=\pmatrix{1&1&0\cr0&z&0\cr0&-z&1\cr}
\qquad &.\cr}
\eqn\matrici
$$
{}From eq.s \matrici\  we may verify that indeed $\a_5$ satisfies the
relation \bifive . Furthermore, we can use \matrici\  to compute the
monodromy generator around $\infty$, which we have disregarded until now.
It can be shown that  the generator $\a_\infty$ can be written by the
following word
$$
\a_\infty=\a_4 \a_2 \a_3 \a_5 \a_1 \a_5
\eqn\parola
$$
and we obtain
$$
\cR(\a_\infty)=\pmatrix{z^2&1+z^2&z(1+z^2)\cr
-z^2&-1-z^2&-z^3\cr
-1&-1&-1-z\cr}
\eqn\ainfi
$$
We note that the eigenvalues of $\cR(\a_\infty)$ are $\{-1,-1,-z\}$
 thus showing the presence
of a singularity at $\infty$ with critical exponent$-{2\over5}$. This can also
be confirmed by the behaviour of the integral \periodo\ for large values
of $a$ and $b$. We find in this case
$$
\o_0{\buildrel {a ,b \to\infty}\over\sim}\oint{{d\tau}\over{(5 a \tau^3+ 5 b
\tau^2)^{2/5}}}
\eqn\quindici
$$
and by the rescaling $a\to \lambda\xi\, ,b\to\lambda\eta$  we find
$$
\o_0\sim\oint{{d\tau}\over{(\xi \tau^3+\eta \tau^2)^{-2/5}}} \ \lambda^
{-2/5}={\it const}\, \lambda^{-2/5}
\eqn\sedici
$$
thus confirming the critical behaviour computed from \ainfi .

\chapter{The Duality Group}

It is known that the full duality group of the moduli space is given not only
by the monodromy group of the periods, $\IGa_M$, but also by the symmetry group
of the defining polynomial $\cW$, $\IGa_\cW$. We now want to show that the
symmetries of the defining polynomial $\cW(\iy;a,b)=0$ give at most a central
extension for the monodromy group $B_5$ acting on the $3$--dimensional basis
of the periods.

It is easily seen that the  transformations leaving invariant $\cW$, are
given by
$$
\cases{
a & $\to\r\, a$\cr
b & $\to\r^{-1}\, b$\cr}
\eqn\duali
$$
with $\r^5=1$ , as they can be undone by the linear coordinate
transformation
$$\cases{
\iy_4 & $\to \r\, \iy_4$\cr
\iy_5 & $\to \r^{-1}\, \iy_5$\cr}
\eqn\undo
$$
Since there
is apparently no other action with this property, we conclude that the duality
group of the superpotential is given by $\IZ_5$.

In order to find the representation $U$ of the transformations \duali\
on the periods, we observe that on any integral, say $\o_{51}$,
the transformation \duali\ can be compensated in the integrand by the map
$$
\tau\to\r^3\tau
\eqn\tautau
$$
\ni On the other hand, choosing $\r=e^{4\pi i/5}$, the transformations
 \duali\ on the $(p,q)$ real
plane correspond to a rotation of an angle $4\pi/5$, mapping the $5$--th
branch of the hypocycloid into the $1$--st, so that $\c_{51}$ is
mapped into $\c_{12}$. Taking into account that $d\tau\to\r^3 d\tau$,
we find
$$
U\, :\,\o_{51}\to e^{2\pi i/5}\,\o_{12}
\eqn\mappa
$$
and analogous relations for cyclically permuted indices.
We now observe that the transformation $\o_{5,1}\to\o_{1,2}$ is realized by
the monodromy operator
$$
Z=\cR(\a_3\a_2\a_1\a_5)=\pmatrix{0&1&0\cr0&0&1\cr z^4&-1+z-z^2&z-1\cr}
\eqn\permu
$$
which corresponds to a generator of the cyclic subgroup $\IZ_5\subset B_5$.
It follows that
$$
U Z\,=\,e^{2\pi i/5}\,\II
\eqn\basta
$$
on any period $\o_{i,i+1}$ and therefore also on the selected basis
$\{\o_{45},\o_{51},\o_{12}\}$. Thus we conclude that, unless there is some
element of $B_5$ represented by $UZ$, the $U$--transformation
gives a central extension of the braid group $B_5$. The above central extension
gives the full duality group of the moduli space of the \CY manifold.

Notice that our result differs from the previously studied one--dimensional
examples, where the full duality group $\IGa$ was given by the semidirect
product $\IGa_M\semi\IGa_\cW$ of the monodromy group of the periods
and the symmetry group of $\cW$, as suggested in \lsw\ . We further remark
that, since the moduli space is $2$--dimensional, we may take as coordinates
the ratios $t_1={{\o_{12}}\over{\o_{51}}},t_2={{\o_{45}}\over{\o_{51}}}$
which correspond to a linear combination of the ``special'' variables of
Special Geometry. Hence, on $t_1,t_2$, the action of the full duality group
is given by a faithful projective $3$--dimensional representation of $B_5$.

Recalling that the full symmetry of the moduli space is given by modding
out by $\IGa$ the local moduli space , we obtain that the geometry of
$\cM$ is given by
$$
\cM={{U(1,2)}\over{U(1)\otimes U(2)}}/{\hat B_5}
\eqn\geometria
$$
where $\hat B_5$ is the previously introduced central extension of $B_5$.

\chapter{Conclusions}
Some comments are in order. We have found a $3$-dimensional representation
of the monodromy group for the three fundamental periods $\o_{4,5} ,
 \o_{5,1} ,\o_{1,2}$ given by the three integrals associated to independent
 loops of the integral
\periodo\ , or, equivalently, to the top solution of the system of
differential equations \linear\ . We know that $B_5$ must act as a group of
discrete isometries on the local moduli space ${{U(1,2)}\over{U(1)\otimes
U(2)}}$ and therefore our matrices should belong to the $U(1,2)$ group.
Infact, it can be shown that the matrices \matrici\  satisfy
$$
{\cR(a_i^\dagger)}\, g\, \cR(a_i)\,=\, g
\eqn\daga
$$
where $g$ is the metric given by
$$
g=\pmatrix{1&{{1-z}\over z}&0\cr{z\over{1-z}}&2+z^2-z^3&z-1\cr0&z-1&1\cr}
\eqn\metrica
$$
Since $g$ has one positive and two negative eigenvalues, indeed $\a_i\in
U(1,2)$. As we have already remarked, there must exist a canonical basis
for the $H_{(3)}$ homology of the \CY , where the direct sum of the
$3$ and ${\overline 3}$ representations of $U(1,2)$ given
by \matrici\  and their complex conjugate take values in $SP(6,\IZ)$, six
being the dimension of $H_{(3)}$. We leave the determination of this change
of basis to forthcoming work.

Let us summarize our results. Starting with the family of curves given in
eq. \pot\  we have been able to compute exactly the duality group of the
periods
associated to $\cW(\iy;a,b)$ by means of some very efficient and powerful
 techniques of algebraic geometry, without resorting to the explicit
computation
of the periods  \eg\  via solutions of the Picard-Fuchs equations. Our method
is
in principle applicable also to more complicated situations where more moduli
are present and/or Yukawa couplings are non--vanishing. In fact, the
computation
of the fundamental group $\pi_1(\CP(N)-L^{N-1};B)$ is always possible in
 virtue of the fundamental theorems of Picard--Severi and Zariski, together
with the Van Kampen relations. The actual construction of the monodromy
group relies however also on the knowledge of the behaviour of the periods
around the singular locus of the defining polynomial. In our case, this
computation was derived from the study of the $1$--dimensional integral,
which simplifies the actual task. It is clear that in general one cannot
 expect that the periods can always be reduced to such one--dimensional
 integrals, and the exam of the leading singularity can be more involved.
Still, it is important to realize that, as mentioned in chapter 3, the
analysis of the singularities can always be done in a systematic way
from the linear system of Picard--Fuchs equations using standard techniques
of fuchsian analysis.

We hope  to extend these techniques for the computation of the duality group
to more complicated examples in  future publications.

\ack{It is a pleasure to thank G. Ponzano for many illuminating discussions
 during the early stages of this work. We also thank P. Candelas, X. De La
Ossa and S. Ferrara for many interesting comments.}

\refout
\end